# Stochastic Single Flux Quantum Neuromorphic Computing using Magnetically Tunable Josephson Junctions

Stephen E. Russek, Christine A. Donnelly, Michael L. Schneider, Burm Baek, Mathew R. Pufall, William H. Rippard, Peter F. Hopkins, Paul D. Dresselhaus, Samuel P. Benz

NIST, Boulder CO

*Abstract*— Single flux quantum (SFQ) circuits form a natural neuromorphic technology with SFQ pulses and superconducting transmission lines simulating action potentials and axons, respectively. Here we present a new component, magnetic Josephson junctions, that have a tunablility and re-configurability that was lacking from previous SFQ neuromorphic circuits. The nanoscale magnetic structure acts as a tunable synaptic constituent that modifies the junction critical current. These circuits can operate near the thermal limit where stochastic firing of the neurons is an essential component of the technology. This technology has the ability to create complex neural systems with greater than $10^{21}$ neural firings per second with approximately 1 W dissipation.

## I. INTRODUCTION

Single flux quantum (SFQ) logic[1-3] relies on voltage pulses generated by $2\pi$ phase slips of the superconducting order parameter across a Josephson junction. These voltage pulses have a time-integrated amplitude given by the flux quantum $\phi_0 = 2.068 \times 10^{-15}$ Vs. Complex logic circuits have been fabricated using this technology and major efforts are ongoing to fabricate a new class of low power SFQ supercomputers.[4] Neuromorphic variations of SFQ logic have been proposed and fabricated.[5-7] These papers have highlighted the natural analogy between SFQ pulse trains and action potentials. However, key components of neural operation, such as neuromorphic synapses that are dynamically reconfigurable and system operation near the thermal limit, have not been addressed. Here, we introduce magnetic Josephson junctions based on reconfigurable magnetic nanoclusters, as a new component for neuromorphic SFQ. Further, we develop stochastic models required to characterize and design SFQ neural circuits which, to achiever ultra-low power, need to operate near the thermal limit. In addition to achieving ultra-low power, operating in the partially stochastic regime has been identified as important for efficient neuromorphic learning systems.[8]

Design tools and state-of-the-art fabrication facilities have been developed for SFQ logic circuits including fabrication on 200 mm wafers, planarized multilayer wiring and high junction uniformity.[9] These capabilities can be leveraged to develop high-complexity very fast neural systems that may be suitable for both application specific computing such as image analysis and, more importantly, for computations that require cognitive processes.

While SFQ logic circuits require a high degree of device uniformity, this constraint is greatly relaxed for neuromorphic circuits. Further, SFQ logic requires precise timing and clock distribution, at frequencies of 10 GHz and above, which can be very challenging, while neuromorphic SFQ is asynchronous (although strongly correlated). Finally, SFQ circuits usually require significant bias currents, typically 100 μA per device, which can lead to a prohibitively large total current when device counts are on the order of $10^9$. Stochastic SFQ works very close to equilibrium and does not require significant bias currents.

## II. STOCHASTIC JOSEPHSON JUNCTIONS

A standard circuit model of a Josephson junction is shown in Fig. 1 along with the equations dictating the junction dynamics.[10] The supercurrent is given by $I_s = I_c \sin(\theta)$, where $\theta$ is the change of the phase of the superconducting order parameter across the junction, $V$ is the voltage across the junction which is proportional to the time derivative of $\theta$, $I_c$ is the critical current, $R_n$ is the normal state resistance of the device, $C$ is the junction capacitance, $I_n$ is the thermal noise term, and $I_b$ is the total current through the junction. Here, $I_n$ is assumed to be random Gaussian noise with rms amplitude of $(4k_BT/R_n\tau)^{0.5}$, where $k_B$ is Boltzmann's constant, $T$ is the device temperature, and $\tau$ is the noise sampling time used in modeling. The resulting dynamics are the same as a driven damped pendulum. The energy barrier, the difference between the energy at $\theta = 0$ and $\theta = \pi$, is given by $I_c\phi_0/2\pi$. For stochastic operation (as shown in the modelling below) we require $I_c\phi_0/k_BT \approx 10$ to $20$, with the exact value being an important circuit parameter. The damping is given by the McCumber parameter $\beta_c = 2\pi I_c R_n^2 C/\phi_0$. Most SFQ circuits require the junctions to be close to critical damping with $\beta_c \approx 1$.

Fig. 2 shows numerical solutions of the dynamical equations for a typical stochastic SFQ junction with an operation temperature of 4 K and a critical current of 0.5 µA. Fig. 2b shows the resting state spiking characteristic of the junction, while Figs. 2a, c show the spiking with small applied bias currents of ± 0.1 µA. Unlike real neural systems, the resting state can generate either positive or negative pulses and the pulse durations are < 100 ps. We see that small bias currents can stimulate strong pulse trains. The bias current can be viewed as integrated inputs from other neural spike chains.

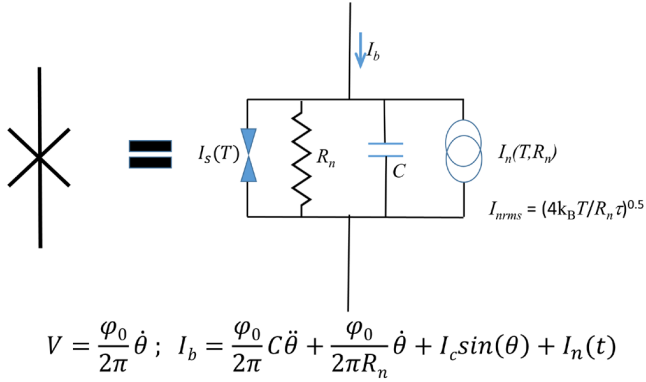

**Figure 1** Circuit model of a Josephson junction, along with the corresponding dynamical equations.

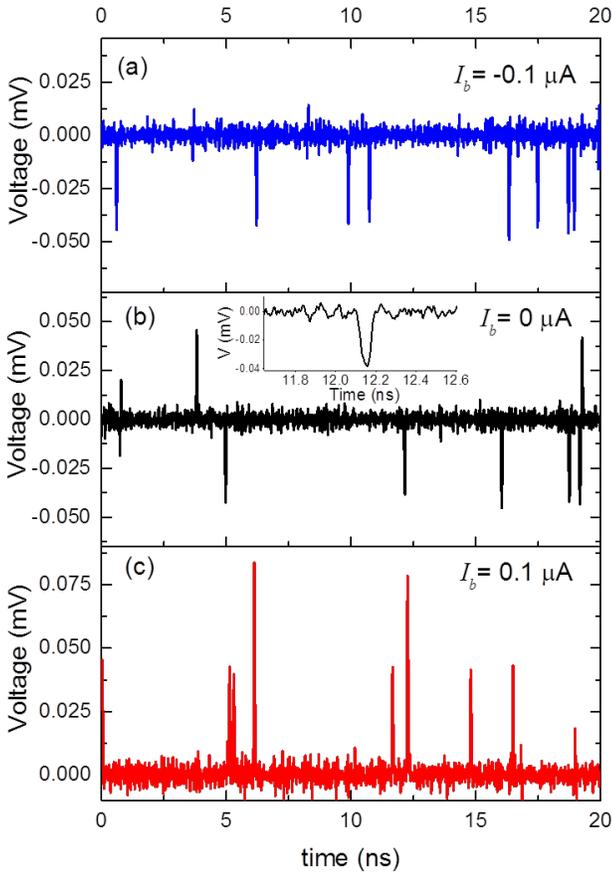

**Figure 2** Simulations of spiking characteristics of a Josephson juction with 0.5 µm diameter, 2 nm thick Si barrier, $I_c$ = 0.5 µA, $R_n$ = 400 Ω, T = 4 K. $I_c\phi_0/k_BT$ = 18: a) -0.1 µA bias current, b) resting state spiking, c) spiking with +0.1 µA bias currents. The circuit bandwidth gives a pulse width near 50 ps.

Fig. 3 shows the spiking characteristic of two otherwise identical junctions with different critical currents, $I_c$ = 0.5 µA and 1.0 µA. Since the spiking rate is exponentially dependent on the critical current, a small change in critical current has a dramatic effect in the spiking characteristics. This type of $I_c$ change is what we hope to dynamically produce through the use of magnetically tunable junctions discussed in the next section.

The spiking rate shown in Fig. 3b is approximately $10^{10}$ single flux quantum per second. The power dissipated for a single active SFQ neuron, the bias current times the average voltage, is therefore $10^{10}$ s$^{-1}I_b\phi_0 \approx 1 \times 10^{-11}$ W. Hence, large scale neural circuits can be fabricated with up to $10^{11}$ neurons and $10^{21}$ firings per second that consume less than 1 W of power (this does not include the refrigeration power).

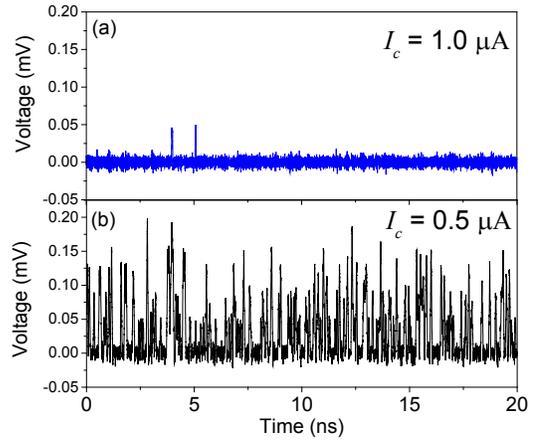

**Figure 3** Simulations of spiking of a 0.5 µm diameter 400 Ω junction at 4.2 K for critical currents of a) 1.0 µA and b) 0.5 µA and a bias current of 0.3 µA. The $I_c$ = 0.5 µA junction shows spikes corresponding to 1, 2 and 3 flux quanta.

### III. MAGNETIC JOSEPHSON JUNCTIONS

Magnetic Josephson junctions have recently received a lot of attention due to their potential use as compact nonvolatile memory elements in SFQ circuits. It has been shown that by changing the magnetic state of a spin valve incorporated into a Josephson junction, one can modulate both the amplitude and phase of the supercurrent.[11-16] For carefully chosen thicknesses of the magnetic layers, the modulation of the critical current can be greater than 100%.[17] Further, it has been shown that in small devices of less than 100 nm diameter, the magnetic state can be changed by current pulses through the device via spin transfer torque.[17, 18]

An existing drawback of the digital magnetic Josephson devices is that large currents are required to switch them. This is due to the fact that the magnetic layers have large moments and that each electron can only transfer $\hbar$ of angular momentum. A large number of electrons are required to transfer the required angular momentum, while only a small percent of the energy of the electrons is used to overcome the magnetic energy barrier. To make the magnetic Josephson junction have an analog $I_c$ variation and to make switching more efficient, we fabricate Josephson junctions with Mn clusters embedded in the Si barrier, as shown in Fig. 4. These Josephson junctions, with the exception of Mn doping, are very similar to the ones used in NIST Josephson voltage standards.[19, 20]

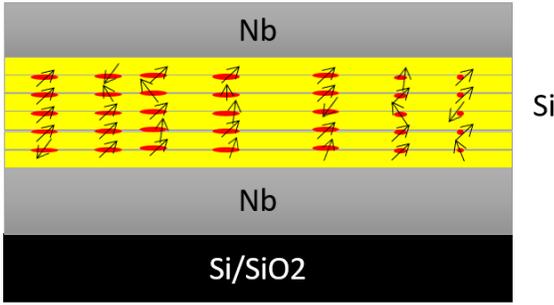

**Figure 4** Schematic of magnetic Josephson junction with Mn clusters embedded in a Si barrier.

The junctions, as seen in Fig. 5, show nearly ideal current-voltage characteristics for overdamped junctions, nearly ideal $I_c$ vs field characteristics, high uniformity as determined by measurements on large arrays, and high quality microwave response. The presence of the magnetic clusters can greatly suppress $I_c$ and it has been shown (to be reported in another publication) that $I_c$, as with other magnetic Josephson junction structures [11, 21-24], depends on the magnetic state. While the critical currents for the junctions shown here are quite large, for smaller junctions with dimensions of 100 nm, which are appropriate for large scale neural networks, the critical currents will be on the order of a few microamps.

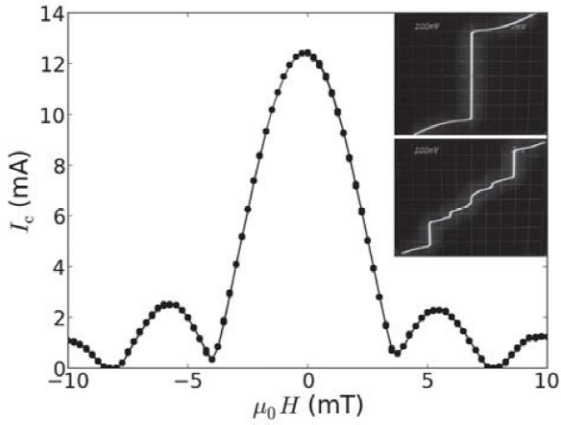

**Figure 5** Critical current vs. magnetic field for a 3 um x 6 μm Mn-doped Josephson junction.[24] The insets show the current-voltage curves for series array of 600 Mn-doped Josephson junctions with and without 4.7 GHz microwaves applied.

The cluster size and switching energies are set by post deposition anneals. The anneals are done in a rapid thermal annealer at 573 K to 723 K for approximately 2 min. Fig. 6 shows typical zero-field-cooled and field-cooled moment vs. temperature data. Below the blocking temperature, indicated by the vertical arrows, the Mn spins can be configured in either random or ordered states. Figure 6a shows a sample that was annealed at 573 K with a blocking temperature of 34 K. Figure 6b shows a sample that was annealed at 723 K with a blocking temperature of 240 K. Figure 6 demonstrates the ability to readily change both the blocking temperature as well as the total moment by varying the annealing temperature. By varying the anneal time and temperature we have been able to vary the blocking temperature, $T_B$, from 10 K to 240 K.

The lower the blocking temperature, the easier the magnetic clusters are to switch, with a tradeoff of loss of thermal stability. The switching energy for a magnetic cluster is given approximately by $E_s \approx \ln(t_m/t_a)\ k_B T_B$, where $t_m$ is a measurement time, and $t_a$ is an attempt time. For a blocking temperature of 30 K, $E_s \approx 70$ meV. The cluster moment can be estimated from Langevin fits to moment versus field curves above the blocking temperature. These fits give the cluster moments of a few Bohr magnetons to several hundred Bohr magnetons, depending on the annealing conditions. A key element of these junctions is that the ratio of the switching energy to the angular momentum of the clusters is close to the ratio of energy to angular momentum of an electron, indicating that efficient energy and angular momentum transfer can occur.

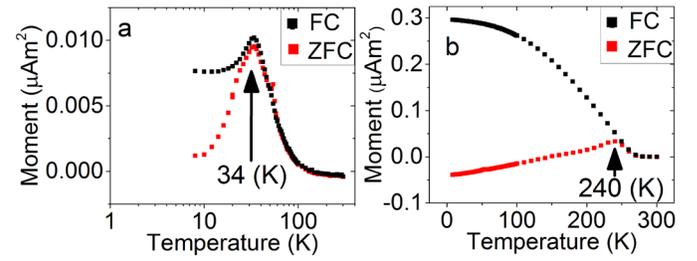

**Figure 6** Zero field cooled (red) and field cooled (black) magnetic moment plots of a Mn-doped Josephson juntion showing hysteretic magnetic states below the blocking temperature, indicated by the vertical arrows.

## IV. MAGNETIC JOSPEHSON JUNCTION DEVICE MODEL

To develop complex neural circuits, it is essential to have a device model that can be incorporated into large scale circuit simulations. We have modified existing Verilog A models of a standard Josephson junction to accommodate an internal magnetic order parameter. Fig. 7 shows the circuit model of a nanocluster magnetic Josephson junction, now with the supercurrent a function of the magnetic order parameter $m$, which in turn is a function of the integrated voltage history. The parameter $m$ is zero when the spins are disordered and 1 when they are completely aligned. Positive applied voltage pulses will drive the system to a disordered state with $m = 0$, while negative pulses will drive the system to $m = 1$.

Here we ignore the dependence of the phase on the magnetic order parameter, $\delta(m)$ = constant, and only allow the critical current to vary. As with the spin valve Josephson junctions, the critical current is maximized in the state with unaligned spins and minimized in the state with spins aligned. The temperature is a critical experimental parameter that can simultaneously scale all critical currents to allow evaluation of the performance with different levels of resting state activity. At present, this model is not validated by experimental data. It represents the simplest dependence of $I_c$ on $m$ to allow simulations to determine the utility of such a device in complex neural circuits.

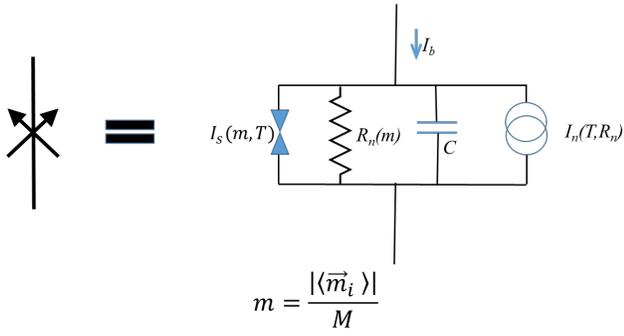

$$m = \frac{|\langle \vec{m}_i \rangle|}{M}$$

$dm = -Vdt/\phi_m$ for $0 < m < 1$;
$dm = 0$ for $0 \geq m \geq 1$
$I_s = I_{co}(m,T)sin(\theta + \delta(m))$
$I_{co}(m,T) = ((1-m)I_{cv} + I_{cm}) * (1 - \frac{T^2}{T_c^2})$

**Figure 7** Circuit model of a nanocluster magnetic Josephson junction along with model equations describing dependence of $I_c$ on the magnetic order parameter.

Given this device model, the performance of devices and circuits can be modeled with SPICE as shown in Fig. 8. This simulation shows similar device response as seen in Fig.1, but now with dynamically variable critical current and done in SPICE, which is capable of extension to large scale circuit simulation. Fig. 8b shows the dynamical response, voltage vs. time, of a magnetic cluster Josephson junction, at $T = 0$ K, for large negative bias, zero bias, and large positive bias. The voltage oscillation amplitude is seen to first decrease with time in the negative current bias regime and then increase with time in the positive current bias regime. This corresponds to a time dependent decrease in the critical current during negative bias (negative input pulses) and an increase in critical current during positive bias. As shown in Fig. 3, this change in critical current can activate a magnetic Josephson junction or deactivate it. An input of negative pulses to a junction will increase $I_c$ causing it to fire more strongly, while a series of positive input pulses will raise $I_c$ preventing future firings.

## SUMMARY

We have introduced nanocluster magnetic Josephson junctions as a potential synaptic component for SFQ neural circuits. We have demonstrated that in the stochastic limit the spiking behavior can be made very sensitive to small changes in critical currents that can be obtained by changing the magnetic structures in magnetic Josephson junctions. We have shown that operating in the stochastic limit can achieve much lower power and eliminate the need for large total bias currents. Finally, we have introduced stochastic circuit models that can be used in large scale SPICE simulations of SFQ neural circuits.

## ACKNOWLEDGMENT

We thank the IARPA C3 program for providing inspiration and guidance for this work.

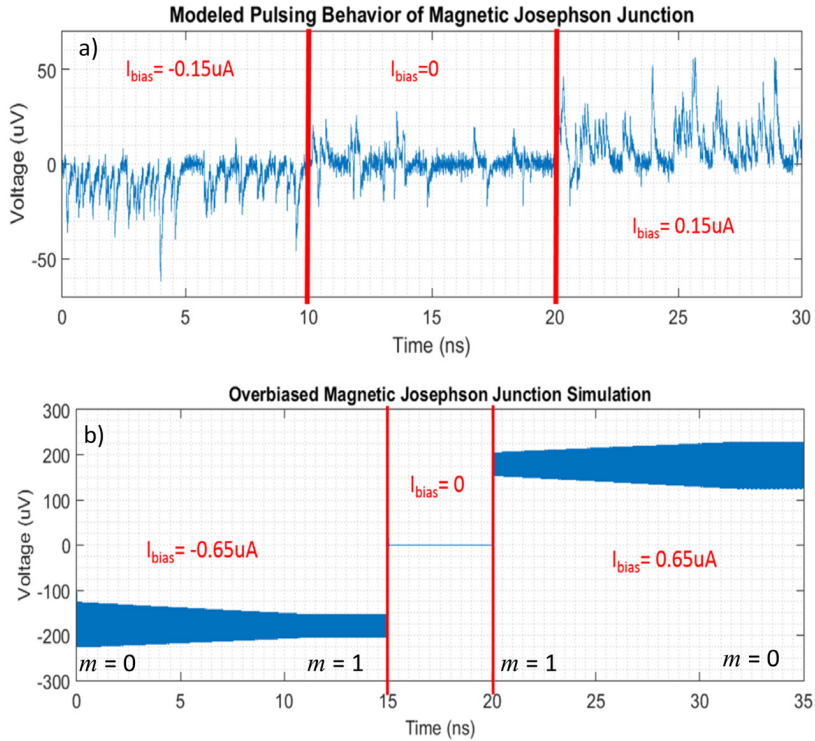

**Figure 8** SPICE simulations showing voltage vs. time of a magnetic Josephson junction, with a critical current of 0.5 μA, when time dependent bias currents are applied. a) Stochastic spiking behavior for a junction at T= 4 K with bias currents of -0.15 μA, 0 μA, and 0.15 μA. b) Oscillation behavior at T = 0 K, with applied bias currents larger than $I_c$ showing dynamic reconfigurablilty of the junction critical current.